\documentclass[a4paper]{amsart}
\usepackage{amsmath}
\usepackage{amsthm}
\usepackage{amssymb}
\usepackage[mathscr]{eucal} 
\newtheorem{Thm}{Theorem}[section]
\newtheorem{Cor}[Thm]{Corollary}
\newtheorem{Prop}[Thm]{Proposition}
\newtheorem{Lemma}[Thm]{Lemma}
\theoremstyle{definition}

\theoremstyle{remark}
\newtheorem{rem}[Thm]{Remark}

\numberwithin{equation}{section}

\newcommand{\al}{\alpha}
\newcommand{\calB}{{\mathcal B}}

\newcommand{\calC}{{\mathcal C}}

\newcommand{\cO}{{\mathcal O}}                                      
\newcommand{\cC}{{\mathcal C}}                                      

\newcommand{\De}{\Delta}

\newcommand{\calF}{{\mathcal F}}
\newcommand{\calG}{{\mathcal G}}

\newcommand{\calH}{{\mathcal H}}
\newcommand{\calK}{{\mathcal K}}

\newcommand{\frakK}{{\mathfrak K}}
\newcommand{\La}{\Lambda}
\newcommand{\calL}{{\mathcal L}}

\newcommand{\calO}{{\mathcal O}}
\newcommand{\bP}{{\mathbb P}}

\newcommand{\calR}{{\mathcal R}}
\newcommand{\R}{{\mathbb R}}
\newcommand{\si}{\sigma}

\newcommand{\calW}{{\mathcal W}}

\newcommand{\Ze}{{\mathbb Z}}

\newcommand{\fLor}{{\mathcal L}}
\newcommand{\Lor}{{\mathcal L}_+^\uparrow}
\newcommand{\Poi}{{\mathcal P}}
\newcommand{\Poif}{{\mathcal P}_+^\uparrow}
\newcommand{\Poip}{{\mathcal P}_+^\downarrow}
\newcommand{\ProPoi}{{\mathcal P}_+}

\DeclareMathOperator*{\Ind}{Ind}

\begin{document}
\author[R. Brunetti, D. Guido, R. Longo]{{\bf 
R. Brunetti$^{\rm (1)}$ 
D. Guido$^{\rm (2)}$ R. Longo$^{\rm (2)}$}\\
\hfill\\
${}^{(1)}$ Dipartimento di Scienze Fisiche\\
           Universit\`{a} di Napoli ``Federico II''\\
           Complesso Univ. Monte S. Angelo\\
           I--80126 Napoli, Italy\\
\hfill\\
${}^{(2)}$ Dipartimento di Matematica    \\
           Universit\`a di Roma ``Tor Vergata''\\
           Via della Ricerca Scientifica 1\\
           I-00133 Roma, Italy} 
	   
\date{}

\title[Modular Localization and Wigner Particles]
{Modular Localization and Wigner Particles}

\dedicatory{\normalsize Dedicated to
Huzihiro Araki on the occasion of his seventieth birthday}

\thanks{e-mails:\ brunetti@na.infn.it, guido@mat.uniroma2.it,
longo@mat.uniroma2.it} 

\thanks{Supported in part by MIUR and
GNAMPA-INDAM} 

\begin{abstract} 
We propose a framework for the free field construction of algebras of
local observables which uses as an input the Bisognano-Wichmann
relations and a representation of the Poincar\'e group on the
one-particle Hilbert space.  The abstract real Hilbert subspace
version of the Tomita-Takesaki theory enables us to bypass some
limitations of the Wigner formalism by introducing an intrinsic
spacetime localization.  Our approach works also for continuous spin
representations to which we associate a net of von Neumann algebras on
spacelike cones with the Reeh-Schlieder property.  The positivity of
the energy in the representation turns out to be equivalent to the
isotony of the net, in the spirit of Borchers theorem.  Our procedure
extends to other spacetimes homogeneous under a group of geometric
transformations as in the case of conformal symmetries and de Sitter
spacetime.
\end{abstract}
\maketitle

\section{Introduction}\label{sec:intro}
Although Quantum Physics represents one of the most innovative and drastic 
conceptual changes of view in modern science, the construction of 
Quantum Mechanics and Quantum Field Theory has been fruitfully 
realized with the guidelines of the ``classical analogue''. This is 
unsatisfactory, beyond the well known difficulties to construct a 
quantum field theory with interaction, if one takes the attitude that
quantum field theory should stand on its own legs \cite{[Haa]}.

One point where the structure is selfconsistently dictated by quantum 
principles is the construction of local observable algebras associated with free 
fields. We may summarize the construction in the following building 
blocks:
\begin{enumerate}
\item The one-particle Hilbert space.
\item Second quantization.
\item Localization.
\end{enumerate}
 
Point 1 is E. Wigner's cornerstone analysis of the irreducible unitary
representations of (the cover of) the Poincar\'e group.  As is well
known, the positive energy representations 
are classified by the mass $m$ and the spin $s$ if $m>0$. 
When $m=0$ the stabilizer of a non-zero point is isomorphic to the
Euclidean group $E(2)$ which is not compact.  Irreducible
representations of the Poincar\'e group induced by finite-dimensional
representations of $E(2)$, namely by representations which are trivial
on the translational part, are labelled by the helicity (a character
on the one-dimensional torus).  Irreducible representations of the
Poincar\'e group induced by infinite-dimensional representations of
$E(2)$ are historically called continuous spin representations
(although properly speaking one should talk of helicity rather than
spin).  Usually one discards such representations because the
corresponding particles have not been experimentally observed so far,
but there is no conceptual a priori reason not to consider them.  As
we will explain below, the analysis in this paper naturally gets into
the consideration of the case of continuous spin too.

Point 2 is well described by E. Nelson's expression: ``First quantization 
is a mystery, but second quantization is a functor''. Segal's 
quantization is indeed an automatic procedure to get Weyl operators on 
the Fock space associated with vectors in the one-particle Hilbert space. In 
particular one gets a von Neumann algebra out of a real Hilbert
subspace of the one-particle space: this is Araki's lattice of 
von Neumann algebras \cite{[Araki],[Arak3]}. 
In this sense free field analysis is basically reduced to one-particle analysis.

In point 3 the basic principle of locality enters. The definition of 
local real Hilbert subspaces, hence of local von Neumann algebras, 
requires however one more step. One possibility is to take 
the functions localized in a region of the configuration spacetime and 
then get the real Hilbert space in the momentum space. That this 
procedure is not entirely intrinsic may be seen from the fact that it 
is not possible to extend it to the case of continuous spin \cite{[Y]}.

The purpose of this note is to show how a net of 
local algebras may be canonically
associated with any positive energy (anti)-unitary representation of the proper
Poincar\'e group. This construction relies on the idea of modular
covariance, namely the identification of some one-parameter subgroups of
the Poincar\'e group with some modular groups constructed via the
Tomita-Takesaki theory. In this way a net of standard subspaces of the
representation space may be canonically defined directly in the Wigner 
one-particle space. Then the second
quantization functor produces the net of local von~Neumann algebras.
Such a net coincides with the one generated by the free Bose field
of mass $m$ and spin $s$ when the corresponding irreducible
representation of the proper Poincar\'e group is considered.
This construction reveals the deep connection between the positivity of
the energy and the isotony property of the net, and reflects the relation
between the cyclicity of the vacuum for the intersection of two wedges and
the existence of a PCT operator in terms of the Tomita modular
conjugations, cf. \cite{gl:95}.
Our analysis is related to \cite{[BL],[Ll]}. 

In other words, the Bisognano-Wichmann theorem tells us what the 
Tomita operator associated with a wedge region $W$ should be. 
Since it is a second quantization operator \cite{[EcOs1]}, 
it is determined by the operator $S_W$ on the one-particle Hilbert 
space $\calH$. 
According to Bisognano-Wichmann 
\begin{equation}\label{S}
S_W=J_W \Delta_W^{1/2}
\end{equation}
is made up by the boosts unitaries $\Delta_W^{it}$ and the PCT
anti-unitary that are canonically associated with the given
(anti)-unitary irreducible representation of the proper Poincar\'e
group.  We may then reverse the point of view and define $S_W$ by
formula (\ref{S}) in terms of the Poincar\'e group representation, hence
define the real subspace
\[
\calK_W\equiv \{\xi: S_W\xi=\xi\}.
\]
This procedure is, of course, general and can be performed for any
unitary representation of the Poincar\'e group, including those with
continuous spin, where the construction of the corresponding Wightman
fields is not possible \cite{[Y]}.

The  von Neumann algebra $\calR(W)$ is then defined by
\[
\calR(W)=\{V(\xi):\xi\in\calK_W\}''\ ,
\]
where $V$ is the representation of the Weyl commutation relations on 
the Fock space over $\calH$.
If $\cO$ is a region of the spacetime obtained as 
intersection of wedges, we may then define
\[
\calR(\cO)\equiv\bigcap_{W\supset\cO}\calR(W)
\]
(intersection over all wedges containing $\cO$). By a classical result
the vacuum vector $\Omega$ is cyclic for $\calR(\cO)$ if $\cO$ is a 
double cone, for any irreducible representation of finite helicity.
By an intrinsic analysis in terms of Poincar\'e group representations, we shall show 
that, in case of continuous spin, $\Omega$ is cyclic for $\calR(\cO)$ if $\cO$ is a 
space-like cone. But Reeh-Schlieder property for double cones is not to 
be expected in this case \cite{[IM]}.

Our analysis extends to spacetimes with a group of symmetries, where a
suitable notion of ``wedge region'' can be defined, in particular to
any such wedge one would associate a one-parameter group of symmetries
and a time-reversing reflection, both giving rise to modular objects
in the unitary representations.  The precise context is explained in
Section \ref{NonMinkowski}, cf. also \cite{BDFS,GLRV} for related 
notions of wedge.  Relevant situations are those given by
the Minkowski spacetime (or the covering of its Dirac-Weyl
compactification) with conformal symmetries, by the circle with
M\"obius transformations, and by the ($d$-dimensional) de Sitter
spacetime with the isometry group $SO(d,1)$.

Preliminary versions of this article have been circulating since a few 
years. The concept of modular localization has then found different 
applications in papers by B. Schroer and collaborators, see 
\cite{[FS]} and references therein.

\section{Basic Preliminaries}
\label{sec:prim}
Let us recall 
some basic geometrical and analytical facts.
The most important geometrical setting we consider is 
Minkowski spacetime, but we shall abstract 
our procedure to extend it to more general spaces and to 
discuss some other examples. 

The Minkowski spacetime is the real manifold
$\R^d \equiv\R\times\R^{d-1}$ of dimension $d\ge 2$, 
equipped with the metric 
\begin{equation*}
\langle x,y\rangle =x^0 y^0 - \sum_{i=1}^{d-1}x^i y^i\ , 
\qquad \forall x,y\in\R^d\ .
\end{equation*}
This makes Minkowski space a Lorentzian manifold and we consider the
time orientation fixed once and for all.  As a result the Minkowski
spacetime is divided into subregions called spacelike, timelike and
lightlike corresponding resp.  to $\langle x,x\rangle< 0$, $\langle
x,x\rangle> 0$, and $\langle x,x\rangle = 0$.

By theorems of Zeeman, the group of diffeomorphisms of the
Minkowski space preserving the causal structure is the semidirect product 
of $\R\times\fLor$ with the translations, where
$\R$ acts as the group of dilations and $\fLor$ is the full 
homogeneous Lorentz group.
On the other hand the group of 
isometries of the Minkowski space is
the Poincar\'e group $\Poi$, the semidirect
product $\fLor\ltimes\R^d$, where $\R^d$ corresponds to the
spacetime translations:
\begin{equation*}
   (\Lambda,a)\circ (\Lambda^\prime,b)=(\Lambda\cdot\Lambda^\prime,
   a+\Lambda\cdot b)
   \ ,\qquad \mbox{with}\,\ \Lambda,\Lambda^\prime\in\fLor,\ 
   a,b\in\R^d \ .
\end{equation*}
The full Poincar\'e group $\Poi$ is simply connected, non connected, non 
compact, and perfect. It admits a splitting into 
connected components 
\begin{equation*}
   \Poi = \Poif\cup\Poip\cup\Poi_{-}^{\uparrow}\cup\Poi_{-}^{\downarrow}\ .
\end{equation*}
where the $\pm$ corresponds to $\mbox{det}(g)=\pm 1$, namely selects 
those transformations which preserve or change the orientation, and 
the up/down arrow corresponds to $\langle x, gx\rangle\gtrless 0$, namely 
selects those transformations which preserve or change the time 
orientation.

We shall be mainly concerned with the {\sl proper} part of the 
Poincar\'e group, i.e. $\ProPoi = \Poif\cup\Poip$.

Let then $\ProPoi\ni g\longrightarrow U(g)$ be a strongly continuous 
(anti-)unitary representation on the Hilbert space $\calH$, i.e., 
\begin{equation*}
   U(g)\,\,\,\mbox{is}\,\,\,
 \begin{cases}
  \text{unitary}& \mbox{if}\,\, g\in\Poif\\
  \text{antiunitary}& \mbox{if}\,\, g\in\Poip
 \end{cases}
\end{equation*}
We select now a particular class of causally complete subregions in
Minkowski spacetime which are left globally invariant by suitable
one-parameter velocity transformations.  It is traditional to call
them {\sl wedge} regions and we denote the set of wedges by $\calW$. 
As usual, $W'$ denotes the causal complement of $W$.
Each wedge is a Poincar\'e transform of the wedge $W_1 =\{x\in\R^d:
x_1 > |x_0|\}$.  It is possible to assign to each wedge a one
parameter group of transformations $\Lambda_{W}$ and a time-reversing
reflection $R_{W}$ satisfying
\begin{itemize}
	\item[{\bf (a)}] {\bf Reflection covariance.} For any
	$W\in\calW$, $R_W$ maps $\calW$ onto $\calW$, $R_{W}(W)=W'$
	and $R_{gW}=gR_{W}g^{-1}$, $g\in \ProPoi$.
	 \item[{\bf (b)}] {\bf $\Lambda$-covariance.} For any $W\in\calW$, 
	 $\Lambda_W(t)$ maps $\calW$ onto $\calW$, $\Lambda_{W}(t)(W)=W$ 
	 and $\Lambda_{gW}(t) = g\Lambda_{W}(t)g^{-1}$, $t\in\R$, $g\in 
	 \Poif$, $\Lambda_{gW}(t) = g\Lambda_{W}(-t)g^{-1}$, $t\in\R$, 
	 $g\in \Poip$.  %
\end{itemize}
Indeed, since the action of $\ProPoi$ is transitive on the family 
$\calW$, it is enough to choose $\Lambda_{W_{1}}$ and $R_{W_{1}}$ to 
determine the whole assignment. Moreover, setting 
$\Poif(W):=\{g\in\Poif : gW=W\}$, properties {\bf (a)} and {\bf (b)} imply 
that $\Lambda_{W}$ is in the center of $\Poif(W)$, while $R_{W}$ 
commutes with $\Poif(W)$.

$\Lambda_{W_{1}}$ is chosen as the (rescaled) boosts preserving $W_{1}$, 
namely
\begin{equation*}
   \La_{W_1} : \R\ni t\to\La_{W_1}(t) = 
\left (\begin{array}{ccccc}
 \cosh(2\pi t)& -\sinh(2\pi t) & 0      & \dots    & 0 \\
 -\sinh(2\pi t)& \cosh(2\pi t) & 0      & \dots    & 0 \\
     0        &       0       & 1      & \dots    & 0  \\
     \vdots   &      \vdots   & \vdots & \ddots   & \vdots \\
     0        &       0       & 0      & \dots    & 1
\end{array}\right ) \in\Lor \ .
\end{equation*}
The element $R_{W_1}$ in $\ProPoi$ is the 
reflection w.r.t. the edge of the wedge $W_1$, and is given by 
\begin{equation*}
   R_{W_{1}}(x_0,x_1,\dots,x_{d-1})=(-x_0,-x_1,x_2,\dots,x_{d-1})\ .
\end{equation*}

Let's fix a unitary representation $U$ of $\ProPoi$ on a Hilbert space 
$\calH$. With $W\in\calW$ a wedge, 
let $H_W$ be the self-adjoint generator of $U(\La_W (t))$ and define
\begin{align*}
   \De_W  &:= \exp (H_W)\\
     J_W  &:= U(R_W).
\end{align*}

\begin{Prop}\label{Facts}
    The following facts hold true:
    \begin{enumerate}
	\item[$(i)$] $\De_W$ is a densely defined, closed, positive 
	non-singular linear operator on $\calH$;
	\item[$(ii)$] $J_W$ is a antiunitary operator on $\calH$ and $J_W^2={\bf 1}$;
	\item[$(iii)$] $J_W \De_W J_W^{-1} =\De_W^{-1}$.
    \end{enumerate}
\end{Prop}

\begin{proof}
    $(i)$ and $(ii)$ are obvious.  Concerning $(iii)$, let us observe
    that $R_W$ commutes with $\La_W (t)$ which implies that $J_W
    \De_W^{it} J_W^{-1} =\De_W^{it}$, but from the anti-unitarity of
    $J_W$ we have that $J_W H_W J_W^{-1}$ $=$ $-H_W$, hence the
    thesis.
\end{proof}     

These properties allow us to introduce and discuss the properties of
the following operator
\begin{equation*}
   S_W := J_W\,\De_W^{1/2}\ :\ \calH\longrightarrow\calH\ ,
\end{equation*}
indeed, denoting by $R$ and $D$ the range and the domain, we have:
\begin{Prop}\label{prop:S-prop} 
$S_W$ is a densely defined, antilinear, closed operator on 
$\calH$ with $R(S_W)=D(S_W)$ and $S_W^2 \subset 1$.
\end{Prop}
\begin{proof}
 Density and closedness follow from the corresponding property of
 $\De_W$ in Proposition \ref{Facts} $(i)$, antilinearity from the antilinearity of
 $J_W$. Now, $R(S_W)\subset D(S_W)\equiv D(\De_W^{1/2})$, indeed by
 Proposition \ref{Facts} $(iii)$ we have that $J_W\De_W^{1/2}x=\De_W^{-1/2}J_W 
 x\in D(\De_W^{1/2})$. But we get immediately that $S_W^2 =J_W\De_W^{1/2}J_W
 \De_W^{1/2}=\De_W^{-1/2}\De_W^{1/2}\subset 1$ and therefore if 
 $x\in D(S_W)$ then $x=S_W(S_W x)\in R(S_W)$, so we can conclude.
\end{proof}
Let us now define real subspaces of $\calH$ associated with any 
$W\in\calW$, $\calK_W =\{h\in D(S_W)\,\, :\,\, S_W h=h\}$.  Recall 
that an $\R$-linear subspace $G$ in $\calH$ is said to be {\sl 
standard} whenever the following holds:
 \begin{align}
  G \cap iG &=\{0\}\ ,\label{S1}\\
  \overline{G +iG}&=\calH\ .\label{S2}
 \end{align}

\begin{Prop}\label{prop:standard} 
Each $\calK_W$ is an $\R$-linear closed and standard subspace in
$\calH$, $S_{W}$ is the Tomita operator of $\calK_{W}$, namely
$D(S_{W})=\calK_{W}+i\calK_{W}$ and $S_{W}(h+ik)=h-ik$,
$h,k\in\calK_{W}$.  In particular we have:
 \begin{align*}
  \De_W^{it} \calK_W &= \calK_W\\
  J_W\calK_W &= \calK_W^\prime \ ,
 \end{align*}
 where  $\calK_W':=\{h\in\calH\,\,:\,\, {\rm Im}(h,k)=0\quad\forall 
  k\in\calK_W\}$ is the symplectic complement of $\calK_W$.
\end{Prop}
 \begin{proof} The $\R$-linearity and subspace property of any 
 $\calK_W$ is obvious.  Note first than any $x\in D(S_W)$ can be 
 written as $x=h+ik$ where $h$ resp.  $k$ have the form
 \begin{equation*}
 h=  \frac{x+S_W x}{2}\ , \qquad\qquad   k= i\ \frac{-i(x-S_W x)}{2}\ .
 \end{equation*}
 By the preceding Proposition both terms belong to $\calK_W$.  Hence 
 $\calK_W + i\calK_W = D(S_W)$ which is dense, so $(\ref{S2})$ is 
 fulfilled, and if $x\in \calK_W\cap i\calK_W$ then $x=S_W x$ and 
 $ix=S_W ix=-i S_W x=-i x$, therefore $x\equiv 0$, and $(\ref{S1})$ 
 holds too.

 The graph norm on $ D(S_W)$ is, for $x=h+ik$ where 
 $h,k\in\calK_W$,
 \begin{align*}
 \| h+ik\|^2_{S_W} & =\|h+ik\|^2 +\|S_W(h+ik)\|^2\\
                   & =\|h+ik\|^2 +\|h-ik\|^2\\
                   & =2(\|h\|^2 +\|k\|^2) .
 \end{align*}
 Therefore $ D(S_W)$ with the graph norm is $\calK_W\oplus i\calK_W$, hence
 the closedness of $\calK_W$ follows from that of $S_W$.
\end{proof}

\begin{Prop} The representation $U$ acts covariantly on the family 
 $\{\calK_{W}:W\in\calW\}$, namely, 
 \begin{equation}\label{xxx}
  U(g)\calK_W =\calK_{gW}\quad g\in\Poif .
 \end{equation}
\end{Prop}
\begin{proof} From properties {\bf(a)} and {\bf(b)} it follows that
 \begin{equation*}
  U(g)\De_W^{it} U(g)^* =\De_{gW}^{it}
 \end{equation*}
 and
 \begin{equation*}
  U(g) J_W U(g)^* =J_{gW}
 \end{equation*}
 which imply that
 \begin{equation*}
  U(g) S_W U(g)^* = S_{gW}
 \end{equation*}
 hence the thesis.
\end{proof}
Note that eq. (\ref{xxx}) holds true also for $g\in\Poip$ due to 
Prop. \ref{prop:standard} and the following theorem.
\begin{Thm}\label{factor}
	Let $U$ be a (anti-) unitary representation of $\ProPoi$ and 
	$W\mapsto\calK_{W}$  the above defined map. Then wedge duality holds, namely
	$$
	\calK_{W'}=\calK_{W}'.
	$$
	Moreover, the following are equivalent:
\begin{itemize}
	\item[$(i)$] The spaces $\calK_{W}$ are factors, namely 
	$\calK_{W}\cap\calK_{W}'=\{0\}$.
	\item [$(ii)$] The representation $U$ does not contain the trivial 
	representation.
	\item [$(iii)$] The net is irreducible, namely
	$$
	\bigcap_{W\in\calW}\calK_{W}=\{0\}.
	$$
\end{itemize}
\end{Thm}
\begin{proof}
	Observe that $S_{\calK'}=S_{\calK}^{*} = J_{\calK}\ 
	\De^{-1/2}_{\calK}$.  Since $R_{W'}=R_{W}$ and $\Lambda_{W'}(t) = 
	\Lambda_{W}(-t)$, we get the first statement.  Let's prove the 
	equivalences.\\
	$(ii)\Rightarrow(i)$.  We have $\calK_{W}\cap\calK_{W'} = 
	\calK_{W}\cap\calK_{W}' = \{x:U(\La_{W}(t))x =x=J_{W}x, \forall 
	t\in\R\}$.  If such a space contains a non-zero $x$ then the 
	matrix coefficient $(x,U(g)x)$ does not vanish at infinity.  By 
	the vanishing of the matrix coefficient theorem for semisimple Lie 
	groups (cf.  e.g. \cite{[Z]}) the representation must admit an 
	invariant vector.  \\
	$(i)\Rightarrow(iii)$.  This follows directly by the first 
	statement. \\
	$(iii)\Rightarrow(ii)$.  Decompose $U|_{\Poif}$ as
	$U^{0}\oplus I$ where $I$ is the trivial representation (with
	some multiplicity) and $U^{0}$ does not contain the trivial
	representation.  The commutation relations between
	$\Delta^{it}$ and $J$ imply that any $J$ decomposes
	accordingly, namely has no anti-diagonal terms.  Hence any
	space $\calK_{W}$ decomposes as
	$\calK_{W}=\calK_{W}^{0}\oplus\calK_{W}^{I}$.  We have
	$U^{I}(\Lambda_{W}(t))=I$, and, given two wedges $W_{1}$,
	$W_{2}$, $J^{I}_{W_{1}}J^{I}_{W_{2}} =
	U^{I}(R_{W_{1}}R_{W_{2}}) = I$, namely $\calK_{W}^{I}$ is
	independent of $W$. Therefore 
	$$
	\bigcap_{W\in\calW}\calK_{W}\supset0\oplus\calK^{I}.
	$$
	Irreducibility implies $\calK^{I}=0$, namely $U=U^{0}$.	
\end{proof}

\begin{rem}\label{rem:PCT}
	Let us note that the construction of the net $\calK_{W}$ requires
	a representation of $\ProPoi$, or, equivalently, a representation
	of $\Poif$ and a PCT operator.
    
	More precisely we need an anti-unitary involution $J$ satisfying
	$JU(g)J=U(RgR)$, for some space-time reflection $R$. Such involution 
	does not necessarily  exist in any representation. 
	However, given a representation $U$ of $\Poif$ on $\calH$, a
	reflection $R$ and an anti-unitary involution $C$ on $\calH$, we
	may set
    $$
	\tilde{U}(g)=
	\left(\begin{matrix}U(g)&0\\0&CU(RgR)C\end{matrix}\right),\quad
	g\in \Poif,\qquad\tilde{U}(R)= \left(\begin{matrix}0&C\\
	C&0\end{matrix}\right).
    $$
	Clearly $\tilde{U}$ gives rise to a (anti)-unitary representation
	of $\ProPoi$ on $\calH\oplus\calH$.

	Moreover, if $U|_{\Poif}$ is irreducible, then the anti-unitary
	involution $U(R_W)$ is unique up to a phase, that does not depends
	on $W$ by covariance.  Hence the family $\{\calK_W\}$ depends only
	on $U|_{\Poif}$ up to unitary equivalence.
    
	It is known (see e.g. \cite{Vara}) that a PCT operator exists for an
	irreducible representation of $\Poif$ (on $\R^{4}$) if and
	only if the representation is induced by a self-conjugate
	representation of the stabilizer of a point, which is always
	the case, except for the finite non-zero helicity
	representations.
\end{rem}

\section{Inclusions of real subspaces and wedges}

\begin{Prop}
\label{iso1} Let $\calK_1$, $\calK_2$ be standard 
subspaces of the Hilbert space $\calH$, and assume that 
$U\calK_1=\calK_2$, with $U$ unitary on $\calH$.  Then 
$\calK_2\subset\calK_1$ {\it iff} $\De_1^{1/2}U^*\subset J_1 
U^*J_1\De_1^{1/2}$.
\end{Prop}

\begin{proof}
The following equivalences hold:
\begin{align*}
\calK_2\subset\calK_1
&\Longleftrightarrow S_2\subset S_1\\
&\Longleftrightarrow UJ_1\De_1^{1/2} U^*\subset 
J_1\De_1^{1/2}\\
&\Longleftrightarrow \De_1^{1/2} U^*\subset 
J_1U^*J_1\De_1^{1/2}.
\end{align*}
\end{proof}
The following theorem is a one-particle analogue of results in 
\cite{[Borc1],[Wi3]}. It is related to the positive energy criterion 
in \cite{[BCL]}.

\begin{Thm}\label{posen} Let $\calK$ be a standard space in the Hilbert
space $\calH$ and $U(a)=e^{iaH}$ a one-parameter group of unitaries on
$\calH$ satisfying
\begin{align}
\De^{it}U(a)\De^{-it}&=U(e^{\mp 2\pi t}a)\label{eqn:dilat}
\\
JU(a)J&=U(-a),\label{eqn:refl}
\end{align}
where $J$ and $\De$ are the modular conjugation and operator 
associated with $\calK$.

The following are equivalent:
\begin{itemize}
\item[$(i)$] $U(a)\calK\subset\calK$
for $a\geq0$;
\item[$(ii)$]  $\pm H$ is positive.
\end{itemize}
\end{Thm}

\begin{proof} By replacing $\calK$ with $\calK'$ it suffices to 
prove the case $H$ positive.
The implication $(i)\Longrightarrow(ii)$ was proved in 
\cite{[Wi3]}. 

$(ii)\Longrightarrow(i)$.
Let us observe that the spectrum of $H$ is acted upon by the group
$\De^{it}$ and by $J$, with $\{0\}$ and $(0,\infty)$ being the
invariant subsets. The corresponding eigenspaces are henceforth
invariant under the action of $\De^{it}$ and $J$, as a consequence
$\calK$ is decomposed in a direct sum of respectively the $H=0$ and
the $H>0$ parts.  Hence the thesis may be proven in the two cases
separately.  When $H=0$ isotony trivially holds.

In the following we assume that $H>0$.
By Proposition \ref{iso1}, together with equation (\ref{eqn:refl}), we get

\begin{equation}\label{eqn:incl}
U(a)\calK\subset\calK \Longleftrightarrow \De^{1/2} U(a)^*\subset
U(a)\De^{1/2} .
\end{equation}

Let $K=\log H$ (it exists since $H>0$), and $M$ the generator of
$\De^{it/2\pi}$. It is easy to see that $e^{i\mu K}$ and $e^{i\lambda M}$
satisfy Weyl's commutation relations, i.e.,
\begin{equation*}
e^{i\lambda M}e^{i\mu K}=e^{i\lambda\mu}e^{i\mu K}e^{i\lambda M} .
\end{equation*}
According to von Neumann's theorem every representation of the Weyl's
commutation relations is equivalent to a multiple of the Heisenberg
representation. Then the relation on the right hand side of (\ref{eqn:incl})
can be checked in just one representation. Because of the equivalence
(\ref{eqn:incl}), it is enough to verify the inclusion
$U(a)\calK\subset\calK$, $a>0$, in one non-trivial representation.

An example is provided by the one-particle space of the conformal
field theory on the line corresponding to lowest weight
representations of $PSL(2,\R)$. Taking $\calK$ as the standard
space associated with the right half-line $(0,\infty)$, and $U(t)$ as
the translations, the relations in the hypothesis are verified
\cite{bglo1}, and the mentioned inclusion of subspaces hold by isotony.
\end{proof}

\begin{rem}
    Condition (\ref{eqn:refl}) is not needed for the implication
    $(i)\Longrightarrow(ii)$ in the above theorem, see \cite{BS}.
    However the condition is necessary for the converse implication.
    Indeed, given $J$, $\Delta$ and $U$ as in the theorem, and
    assuming positivity of the generator of $U(a)$, one may choose a
    unitary $V$ which commutes with $\Delta$, anticommutes with $J$
    and does not commute with $U(a)$, e.g. $V=(\Delta+i)
    (\Delta-i)^{-1}$, and then replace $J$ with $VJ$, the space
    $\calK$ being redefined accordingly.  Now property $(i)$ in the
    theorem above cannot hold, since, by the result of Borchers
    \cite{[Borc1]}, it would imply condition (\ref{eqn:refl}) for the
    new $J$, against the hypothesis.
\end{rem}
Let us denote by $H$ the cone in the Lie algebra of $\Poif$ consisting
of the generators of future-pointing light-like or time-like
translations.  As is known, a unitary representation of $\Poif$ has
positive energy if the corresponding self-adjoint generators are
positive.  Given two wedges $W_0\subset W$, we shall say that $W_0$ is
positively included in $W$ whenever $W_{0}$ can be obtained by $W$ via
a suitable translation $\exp(a_{0}h)$, $a_{0}\geq0$, such that $\pm h\in H$, where we 
denoted by $\exp$ the exponential map from the Lie algebra to the Lie 
group, and
\begin{equation*}
\begin{align*}
\Lambda_W(t) \exp(ah) \Lambda_W(-t)&=\exp(e^{\mp 2\pi t}ah)
\\
R_{W}\exp(ah)R_{W}&=\exp(-ah)
\end{align*}
\qquad a,t\in\mathbb R\ .
\end{equation*}

The following is a well known geometric fact:
\begin{itemize}
    \item[{\bf (c)}] {\bf Positive inclusion.} Any inclusion of wedges is
    the composition of finitely many positive inclusions.
\end{itemize}

\begin{Thm}\label{isotony} Let $U$ be a (anti-)unitary 
	representation of $\ProPoi$, $W_1\subset W_2$ wedges.  Then 
	$\calK_{W_1}\subset\calK_{W_2}$ {\it iff} $U$ is a positive energy 
	representation.
\end{Thm}

\begin{proof}
    Follows immediately from {\bf (c)} and  Theorem \ref{posen}.
\end{proof}

Since causally complete convex regions are intersections of wedges, 
the map $W\to\calK_W$ extends to causally complete, convex regions 
$\cC$ via
\begin{equation}\label{diamonds}
	\calK_\calC =\bigcap_{W\supset \calC}\calK_W ,
\end{equation}
and to general causally complete regions via
\begin{equation}\label{NonConvDiam}
	\calK_\calO =\bigvee_{\calC\subset\calO }\calK_{\calC} ,
\end{equation}
where $\calC$ are convex and causally complete.  Let us observe that
isotony for wedges implies that equation (\ref{diamonds}) is
consistent with the original definition of $\calK_{W}$.

Denote by $\frakK$ the family of all convex causally complete regions. 
Let us point out he following fact (see e.g. \cite{ThWi}):

\begin{itemize}
    \item[{\bf (d)}] {\bf Wedge separation.} For any space-like separated
    $\calO_{1},\calO_{2}\in\frakK$ there exists a wedge $W$ such that
    $\calO_{1}\subset W$ and $\calO_{2}\subset W'$.
\end{itemize}

\begin{Cor}\label{HaagDual} 
    Let $U$ be a positive energy representation of $\ProPoi$.  Then
    the map $\cO\longrightarrow\calK_\cO$ is a local Poincar\'e
    covariant net of real vector spaces, i.e., isotony holds and if
    $\cO_1\subset \cO_2^\prime$ then $\calK_{\cO_1}\subset
    \calK_{\cO_2}^\prime$.  If $\cO$ is a convex causally complete
    region then Haag duality holds, namely
    $\calK_{\cO^\prime}=\calK_{\cO}^\prime$.
\end{Cor}

\begin{proof} 
	The first part of the statement holds by definition.
	\\
	Let us fix $\cO_0\in\frakK$.  If $\calO_{0}$ is a wedge, then
	duality has been proved in Theorem \ref{factor}.  If
	$\calO_{0}$ is not a wedge, then its space-like complement is
	not convex, hence, by {\bf (d)}, we have the following chain
	of identities:
	\begin{equation*}
		\calK_{\cO_0^\prime}^\prime = ( \bigvee_{\substack{ 
		\cO\subset\cO_0^\prime \\
		\cO\in\frakK}
		} 
		\calK_{\cO}
		)^\prime 
		=( \bigvee_{\substack{
		W^\prime\subset\cO_0^\prime \\ 
		W\in\calW}
		}
		\calK_{W^\prime} 
		)^\prime 
		= \bigcap_{\substack{
		W^\prime\subset\cO_0^\prime \\ 
		W\in\calW}
		}
		\calK_{W^\prime}^\prime 
		=  \bigcap_{\substack{
		W\supset\cO_0 \\ 
		W\in\calW}
		}
		\calK_W = \calK_{\cO_0}\ .
	\end{equation*}
	Therefore Haag duality holds.
\end{proof}
\begin{rem} 
	$(1)$ A net of von~Neumann algebras may be obtained via second 
	quantization:
	\begin{equation*}
		\calR(\cO) =\{V(h):h\in\calK_{\cO}\}''
	\end{equation*}
	where $V(h)$ are the Weyl unitaries on the Bosonic Fock space
	$e^{\calH}$.  Weyl unitaries may be defined via
	\begin{align*}
	    V(h)e^0&=e^{-\frac{1}{4}\|h\|^2}
	    e^{\frac{i}{\sqrt2}h},\quad h\in\calH\\
	    V(h)V(k)&=e^{-\frac{i}{2}{\rm Im}(h,k)}V(h+k)\quad 
	    h,k\in\calH
	\end{align*}
	where the {\it coherent} vectors $e^h$ are defined by
	$e^h=\bigoplus^\infty_{n=0}\frac{h^{\otimes n}}{{\sqrt{n!}}}$.
	Coherent vectors turn out to form a total set in $e^\calH$ (see e.g. 
	\cite{Guic} 
	p.32), hence the $V(h)$'s are well defined unitaries.
	The standard property of $\calK_{\cO}$ is equivalent to the 
	Reeh-Schlieder property for $\calR(\cO)$ (cf.  \cite{[Araki],[EcOs1],[LRT]}).

	$(2)$ If $U$ is the irreducible representation of mass $m$ and 
	spin $s$ the map $\cO\longrightarrow \calR(\cO)$ gives the 
	net of local observable algebras for the free field of mass $m$ 
	and spin $s$.  In fact, for these nets the one-particle 
	version of the Bisognano-Wichmann theorem holds, i.e.,
	\begin{align*}
		J_W &= U(R_W)\\
		\De_W^{it} &= U(\La_W (t))
	\end{align*}
	where $J_W$ and $\De_W$ are the Tomita operators of the real space 
	$\calK_W$ of vectors localized in $W$.  This means that $\calK_W$ 
	is effectively reconstructed in terms of the representation $U$.  
	Moreover, it was shown by Araki \cite{[Arak3]} that the map 
	$\cO\longrightarrow \calK_\cO$ is an isomorphism of complemented 
	lattices
	\begin{equation*}
		(\cap, \cup, \mbox{space-like complement})\rightleftharpoons
		( \cap, \vee, \mbox{symplectic complement})
	\end{equation*}
	if $\cO$ is connected, causally complete, with piecewise $C^1$ 
	boundary. This shows that $\calK_{\calO}$ is also reconstructed in 
	terms of the representation $U$.
\end{rem}
 Three questions arise for the subspaces of the described net 
 $\calO\to\calK_{\calO}$: the standard property, the 
 III$_{1}$ factor property (see \cite {[Arak4]}), namely the 
 fact that the corresponding 
 second quantization algebra is a type III$_{1}$ factor,  
  and the intersection property (for convex causally complete regions), namely
\begin{equation}\label{IntProp}
	\calC=\bigcap_{i\in I}W_{i}\Rightarrow
	\calK_{\calC}=\bigcap_{i\in I}\calK_{W_{i}}.
\end{equation}
 When wedge regions are concerned, we proved the standard property,
 the intersection property and the factor property for irreducible
 nets.  The III$_{1}$ factor property (for irreducible nets) and the
 other properties are proved for space like cones in Section~4.
	
\section{Intersections and cyclicity}

	\begin{Prop} Let $\calK_{j}$, $j\in {\mathcal J}$, a family of 
	standard subspaces of a Hilbert space $\calH$, $o$ a distinguished 
	element of ${\mathcal J}$.  Then $\bigcap_{j\in {\mathcal 
	J}}\calK_{j}$ is standard if and only if the space
	\begin{equation}\label{eq:subu}
		\{x\in\calH:x\in D(S_{j}S_{o})\ \&\ S_{j}S_{o}x=x,\forall 
		j\in {\mathcal J}\}
	\end{equation}
	is dense.
\end{Prop}

\begin{proof}
	Since $\calK_{o}$ is standard, $\cap_{j\in {\mathcal 
	J}}\calK_{j}$ is standard if and only if $\cap_{j\in {\mathcal 
	J}}\calK_{j} +i\cap_{j\in {\mathcal J}}\calK_{j}$ is dense.  We 
	contend that the last subspace can be equivalently written as the 
	expression in (\ref{eq:subu}).
	
	Indeed, if $x\in \calK_{j}$ for any $j\in {\mathcal J}$, then $S_{j} 
	x=S_{o}x=x$ for any $j\in {\mathcal J}$.  Since range and domain of the $S$ 
	operators coincide, $S_{o}x$ belongs to the domain of $S_{j}$ and 
	$S_{j}S_{o}x=x$, $j\in {\mathcal J}$.  Hence $x$ belongs to the space in 
	(\ref{eq:subu}).  Such a space being complex linear, it contains 
	also $i\cap_{j\in {\mathcal J}}\calK_{j}$.  
	
	Conversely, if 
	$x\in D(S_{j}S_{o})$ and $S_{j}S_{o}x=x$ $\forall j\in {\mathcal J}$, 
	then, $\forall j\in {\mathcal J}$, $x\in D(S_{j})$, hence it can be 
	written as $x=h_{j}+ik_{j}$ with $h_{j},k_{j}\in \calK_{j}$, and 
	$S_{o}x=S_{j}x$.

	Therefore we get $S_{o}x=S_{j}x=S_{j}(h_{j}+ik_{j})=h_{j}-ik_{j}$, 
	hence $\frac{1}{2}(x+S_{o}x)=h_{j}$ and $\frac{1}{2i} 
	(x-S_{o}x)=k_{j}$, namely $h_{j}$ and $k_{j}$ are independent of 
	$j$ and belong to $\cap_{j\in {\mathcal J}}\calK_{j}$.
\end{proof}
Recalling the definition in equation (\ref{diamonds}), we get the 
following.

\begin{Prop}\label{cycequiv} Let $U$ be a (anti-)unitary 
	representation of $\ProPoi$ on the Hilbert space $\calH$, $\calC$ 
	a convex, causally complete region, $W$ a wedge containing $\calC$, 
	and $\calG(\calC)=\{g\in\Poif:gW\supset\calC\}$.  Then 
	$\calK_\calC$ is standard {\it iff}, denoting by $T(g)$ the 
	operator $\De^{-1/2}U(Rg^{-1}R)\De^{1/2}$, $g\in\Poif$,
	\begin{equation}\label{eq:cycequiv}
		\{x\in\calH:x\in D(T(g))\ \&\ T(g)x=U(g^{-1})x, \forall 
		g\in\calG(\calC)\}
	\end{equation} 
	is dense, where $\De$ and $R$ refer to the wedge $W$.  
	\\
	Also, given two subsets $\calG_{1},\calG_{2}$ of $\Poif$,
	$\cap_{\calG_{1}}\calK_{gW} = \cap_{\calG_{2}}\calK_{gW}$ {\it
	iff}
	\begin{multline}\label{intprop}
		\{x\in\calH:x\in D(T(g))\ \&\ T(g)x=U(g^{-1})x, \forall 
		g\in\calG_{1}\}\\=
		\{x\in\calH:x\in D(T(g))\ \&\ T(g)x=U(g^{-1})x, \forall 
		g\in\calG_{2}\}.
	\end{multline} 
\end{Prop} 
\begin{proof}
	Since the action of $\Poif$ is transitive on the wedges, the first statement
	immediately follows by the previous proposition. The second statement 
	follows by the proof of the previous proposition.
\end{proof}
Now we may tackle the main questions concerning convex, causally
complete regions in the Minkowski space in this approach, namely the
standard property (\ref{S1}), (\ref{S2}), the III$_{1}$ factor
property and intersection property (\ref{IntProp}).

Since local algebras (and local subspaces) are not defined in terms of 
local fields, the classical Reeh-Schlieder argument does not apply.  
However, Proposition \ref{cycequiv} shows that the standardness for a 
given region (or family of regions) is a property of the 
representation $U$, hence group theoretic techniques may be applied.

Intersection property instead has to do with the definition in 
\ref{diamonds}.  Though the local space of a given convex causally 
complete region $\calC$ is defined as the intersection of the spaces 
of all wedges containing it, just a few of them may be enough to 
determine $\calC$.  Would the corresponding intersection of local 
spaces give rise to the same space?  Again, because of the absence of 
local fields, the answer is not trivial, and the group theoretic 
approach may do the job.

\begin{Lemma}\label{decomposition}
	Let $U$ be a (anti-)unitary positive energy representation of 
	$\ProPoi$ on the Hilbert space $\calH$, $\calC$ a convex, causally 
	complete region.  Assume that the representation $U$ (restricted to 
	$\Poif$) decomposes as $\int^{\oplus}U_{\lambda}d\mu(\lambda)$.  Then 
	$\calK^{U}_{\calC}$ is standard if and only if 
	$\calK^{U_{\lambda}}_{\calC}$ is standard for $\mu$-almost all 
	$\lambda$.

	Given $W_{j}$, $j\in {\mathcal J}$, such that $\calC =
	\bigcap_{j\in {\mathcal J}}W_{j}$, then $\calK^{U}_{\calC} =
	\bigcap_{j\in {\mathcal J}} \calK^{U}_{W_{j}}$ if and only if
	$\calK^{U_{\lambda}}_{\calC}=\bigcap_{j\in {\mathcal J}}
	\calK^{U_{\lambda}}_{W_{j}}$ for $\mu$-almost all $\lambda$.
\end{Lemma}

\begin{proof}
	By Proposition \ref{cycequiv}, both properties depend only on
	$U|_{\Poif}$.  The thesis follows by equations (\ref{eq:cycequiv}), 
	(\ref{intprop}).
\end{proof}

\begin{Thm}
	Let $U$ be a (anti-)unitary positive energy representation of 
	$\ProPoi$ on the Hilbert space $\calH$, $\calC$ a spacelike cone.  
	Then the standard property and the intersection property hold.  If 
	$U$ does not contain the trivial representation, then the type 
	III$_{1}$ factor property holds too.
\end{Thm}

\begin{proof}
	Let us prove the standard property.  Clearly we may assume that 
	the vertex of the space-like cone lies at the origin of the 
	coordinates.  
	
	Lemma \ref{decomposition} shows that is enough to check the 
	density of the space in (\ref{eq:cycequiv}) for all the 
	irreducible positive energy representations.  Since this property 
	is known for the positive mass representations and for the zero 
	mass, finite helicity representations, we only have to verify it 
	for the so called continuous spin representations.
	
	Let us now denote by $\calF(\calC)$ the set of wedges containing 
	$\calC$.  Given a wedge in $\calF(\calC)$, we may consider the 
	family of wedges parallel to the given one and still belonging to 
	$\calF(\calC)$.  The intersection of all such wedges is clearly a 
	wedge in $\calF(\calC)$ whose edge contains the vertex of $\calC$, 
	namely the origin.  Because of isotony (Theorem \ref{isotony}),
	$$
	\bigcap_{W\in\calF(\calC)}\calK_{W}
	=\bigcap_{W\in\calF_{0}(\calC)}\calK_{W},
	$$
	where $\calF_{0}(\calC)$ denotes the subset consisting of wedges 
	whose edge contains the origin.
	
	Then, fixing a wedge $W$ in $\calF_{0}(\calC)$ and setting 
	$\calG_{0}(\calC)=\{g\in\ProPoi:gW\in\calF_{0}(\calC)\}$, the complex 
	span of the space $\cap_{W\in\calF_{0}(\calC)}\calK_{W}$ is given by
	\begin{equation}
		\{x\in\calH:x\in D(T(g))\ \&\ T(g)x=U(g^{-1})x, \forall 
		g\in\calG_{0}(\calC)\},
	\end{equation} 
	namely only the Lorentz subgroup is involved.  Therefore the 
	standard property has only to be checked on the restriction to the 
	Lorentz group of the given continuous spin representation $U$.  
	Theorem \ref{mgives0} concludes the proof.

	Let us now prove the intersection property.  Again by isotony,
	we may restrict to the intersection of wedges whose edge
	contains the origin.  If $\calG_{1}$, $\calG_{2}$ are two
	subsets of $\Lor$ such that $\cap_{g\in\calG_{1}}gW =
	\cap_{g\in\calG_{2}}gW=\calC$, then the equality
	$\bigcap_{g\in\calG_{1}}\calK_{gW} =
	\bigcap_{g\in\calG_{2}}\calK_{gW}$ is equivalent to relation
	(\ref{intprop}).  Then the proof goes on as for the previous
	case.

	We finally prove the III$_{1}$ factor property.  It has been
	proved in \cite{[FiGu2]} that if 1 is in the spectrum of
	$\De$, but not in the point spectrum, then the second
	quantization algebra is a type III$_{1}$ factor.  Clearly the
	property $1\in\sigma(\De)\setminus\sigma_{p}(\De)$ is stable
	under direct sums and quasi-equivalence.  Then, by the proof
	of Theorem \ref{mgives0}, it is enough to show this property
	for the finite spin representations.  Indeed this shows the
	property for the regular representation of $\Lor$, hence for
	the restriction to $\Lor$ of the continuous spin
	representations of $\Poif$, since they are quasi-equivalent to
	the regular representation.
	
	Now we follow \cite{[FiGu1]}, where it is shown (Theorem 3.6) that 
	$\De$ can be written as a functional calculus of a selfadjoint 
	operator $B$ via the function $\frac{t+1}{t-1}$, showing in 
	particular that $1\not\in\sigma_{p}(\De)$.  Moreover, using the 
	explicit formula for $B$, one concludes that $B$ is unbounded, 
	hence 1 is in the spectrum of $\De$.
\end{proof}
Now we prove the standard property for light-like strips, namely for 
regions given by $\overline{W}\cap \overline{W'+a}$, where $a$ is a lightlike vector 
parallel to $W$, namely such that $W+a\subset W$.  Such property 
is motivated by the proof of the spin and statistics property for 
spacetimes with bifurcated Killing horizon given in \cite{GLRV}, 
Section 4.2.

\begin{Thm}\label{StandardLightLike}
	Let $W$ and $a$ be as above and assume the spacetime dimension is $d\ne2$.  
	For any positive energy (anti-)unitary representation of 
	$\ProPoi$, $\calK_{\overline{W}\cap\overline{W'+a}}$ is standard.
\end{Thm}

\begin{proof}
	Clearly any wedge containing $L=\overline{W}\cap\overline{W'+a}$ 
	either contains $W$ or contains $W'+a$.  Then, by isotony (Theorem 
	\ref{isotony}),
	$$
	\bigcap_{W\supset L}\calK_{W}=\calK\cap\calK_{W'+a}.
	$$
	Let us assume for the moment that $U$ is the trivial 
	representation.  Then translations and boosts act trivially, 
	namely $\calK_{W'+a}=\calK_{W'}=\calK_{W}$, since 
	$S_{W'}=S_{W}^{*}=S_{W}$.  Therefore we may assume that $U$ does 
	not contain the trivial representation, namely $U$ does not have 
	invariant vectors.  Since $d\ne2$, the vanishing of the matrix 
	coefficient theorem applies (cf.\ e.g.\ \cite{[Z]}, Proposition 
	2.3.5), hence the spectrum of the generator of any light-like 
	translation is strictly positive, i.e. zero is not an eigenvalue.

	As explained before, the standard property is equivalent to the 
	density of the space
	\begin{equation}\label{cyc}
		\{x\in D(\De^{1/2}U(\tau((a))\De^{1/2}) : 
		U(\tau((a))\De^{1/2}U(\tau((a))\De^{1/2}x=x\},
	\end{equation}
	where $\tau(a)$ denotes the translation by $a$.  This property 
	clearly depends only on the restriction of the representation of 
	the Poincar\'e group to the subgroup $\bP_1$ generated by boosts 
	and light-like translations with strictly positive generator 
	(relative to the wedge $W$).  As the logarithm of the generator of 
	translations and the generator of the boosts give rise to (and are 
	determined by) a representation of the CCR in one dimension, the 
	strictly positive energy representations of $\bP_1$ have a simple 
	structure: they are always a multiple of the unique irreducible 
	representation.  Therefore the density of the space in eqn.  
	(\ref{cyc}) holds either always or never, and hence can be checked 
	in the irreducible case.  But this is the case of the current 
	algebra on the circle, where cyclicity holds by conformal 
	covariance.
\end{proof}	

Now we show that some form of the intersection property holds for 
double cones too.

 Let $\calC$ be a diamond generated by a relatively open convex 
 subregion $\Omega$ of some space-like hyperplane $\calG$.  For any 
 $\xi\in\partial\Omega$, let us consider the family $\calF(\xi)$ of 
 the half-spaces in $\calG$ tangent to $\Omega$ at $\xi$, namely the 
 half-spaces containing $\Omega$ and whose boundary contains $\xi$.  
 Being parametrized by the normal vectors at $\xi$, they have a linear 
 structure, and clearly form a closed convex set.  Let us denote by 
 $\calF_{*}(\xi)$ its extreme points, and by $\calF_{*}(\Omega)$ the 
 union $\bigcup_{x\in\partial\Omega}\calF_{*}(\xi)$.  Clearly 
 $\Omega=\bigcap_{h\in\calF_{*}(\Omega)}h$.  We shall call 
 $\calF_{*}(\Omega)$ the minimal family for $\Omega$.  Analogously, 
 denoting with $W_h$ the wedge generated by the space-like half-space 
 $h$, we shall call $\calF_{*}(\calC) = \{W_h : h\in\calF_{*} 
 (\Omega)\}$ the minimal family for $\calC$.  Clearly when $\calC$ is 
 the intersection of a finite number of wedges $W_{i}$, the minimal 
 family $\calF_{*}(\calC)$ consists only of (some) $W_{i}$.
 
\begin{Thm}
	Let $\calC$ be a diamond generated by a relatively open convex 
	subregion $\Omega$ of some space-like hyperplane $\calG$, 
	$\calF_{*}(\calC)$ its minimal family.  Then
	$$
	\calK_{\calC}=\bigcap_{W\in \calF_{*}(\calC)}\calK_{W}.
	$$
\end{Thm}
 
\begin{proof}
	Let $W$ be a wedge containing $\calC$.  Then 
	$W\cap\calG\supset\Omega$.  Since $W\cap\calG$ is a cone given by 
	the intersection of (at most) two half spaces $h_{1}$, $h_{2}$ of 
	$\calG$, then, by the intersection property for space-like cones, 
	one gets $\calK_{W}\supset\calK_{W_{h_1}\cap 
	W_{h_2}}\supset\calK_{\calC}$ and 
	$\calK_{W_{h_1}}\cap\calK_{W_{h_2}} =\calK_{W_{h_1} \cap 
	W_{h_2}}$.  Therefore
	$$
	\calK_{\calC}=\bigcap_{h\in \calF(\Omega)}\calK_{W_h}.
	$$
	Then, again by the intersection property for space-like cones, for 
	any point $\xi\in\partial\Omega$, we may replace 
	$\cap_{h\in\calF(\xi)}\calK_{W_h}$ with 
	$\cap_{h\in\calF_{*}(\xi)}\calK_{W_h}$, since 
	$\cap_{h\in\calF(\xi)}W_h$ is a spacelike cone, and the proof is 
	completed.
\end{proof}

\begin{Thm}\label{cor:correspondence}
	The following pair of classes can be put in one-to-one 
	correspondence: 
	\begin{itemize}
	\item[$(i)$] Positive energy representations of 
	$\ProPoi$.  
	\item[$(ii)$] Local nets of closed real vector spaces on $\frakK$
	satisfying modular covariance, namely $\De_W^{it}\calK_{\calO} = 
	\calK_{\Lambda_{W}(t)\calO}$, and standard property for the 
	space-like cones.
	\end{itemize}
\end{Thm}

\begin{proof}
	The map from $(i)$ to $(ii)$ has been illustrated above.  The 
	inverse map has been constructed in \cite{bgl:93}, getting a 
	representation of the universal covering of $\Poif$.  It has been 
	shown in \cite{gl:95} that such representation is indeed a 
	representation of $\Poif$, and extends to a representation of 
	$\ProPoi$.
\end{proof}

\begin{rem}
	Let $U$ be a unitary representation of $\Poif$ on a Hilbert space 
	$\mathcal H$ which is finite direct sum of irreducible 
	representations each with strictly positive mass.  As recently 
	shown in \cite{[M]}, if $\calF: W\in\calW\to\calF_W$ 
	is a net of standard real subspaces of $\calH$ and $U$ acts 
	covariantly on $\calF$, namely $U(g)\calF_W=\calF_{gW}$, then 
	$\calF_W$ is the standard subspace associated with $W$ and $U$.
\end{rem}

\section{Free nets on different spacetimes}\label{NonMinkowski}

In this section we discuss various
extensions of the previous construction to different spacetimes.
We begin with a general setting.

Let $M$ be a globally hyperbolic spacetime, $G$ a (Lie) group of 
transformations acting on it (e.g. isometries, or conformal 
transformations), $G_{+}$ the subgroup of orientation preserving 
transformations, $G^{\uparrow}$ the subgroup of time-preserving 
transformations, $G_{+}^{\uparrow}$ their intersection.

Assume it is possible to choose a triple $(\calW,R,\Lambda)$ where
$\calW$ is a family of open, causally complete subregions, called
wedges, stable under the action of $G_{+}$, $R:W\to R_W$ is a map from
$\calW$ to time-reversing reflections in $G_{+}$,
$\Lambda:W\to\Lambda_W$ is a map from $\calW$ to one-parameter
subgroups of $G_{+}^{\uparrow}$ satisfying the following properties:

\begin{itemize}
	\item[{\bf (a)}] {\bf Reflection covariance.} For any
	$W\in\calW$, $R_W$ maps $\calW$ onto $\calW$, $R_{W}(W)=W'$
	and $R_{gW}=gR_{W}g^{-1}$, $g\in G_{+}$.
	 \item[{\bf (b)}] {\bf $\Lambda$-covariance.} For any $W\in\calW$, 
	 $\Lambda_W(t)$ maps $\calW$ onto $\calW$, $\Lambda_{W}(t)(W)=W$ 
	 and $\Lambda_{gW}(t) = g\Lambda_{W}(t)g^{-1}$, $t\in\R$, $g\in 
	 G_{+}^{\uparrow}$, $\Lambda_{gW}(t) = g\Lambda_{W}(-t)g^{-1}$, 
	 $t\in\R$, $g\in G_{+}^{\downarrow}$.  %
\end{itemize}

\begin{rem}
	Properties {\bf (a)} and {\bf (b)} imply that $R_{W'}=R_{W}$ and
	$\Lambda_{W'}(t)=\Lambda_{W}(-t)$.  Moreover, if $gW=W$, then $g$
	commutes with $\Lambda_{W}$ and $R_{W}$, namely $\Lambda_{W}$
	belongs to the center of the stabilizer
	$G_{+}^{\uparrow}(W)=\{g\in G_{+}^{\uparrow}:gW=W\}$ and $R_{W}$
	commutes with $G_{+}^{\uparrow}(W)$.

	If $G_{+}^{\uparrow}$ acts transitively on $\calW$, then the
	assignments $W\to\Lambda_{W}$, $W\to R_{W}$ are determined by the
	choice of a one parameter subgroup in the center of the stabilizer
	of one wedge $W_{0}$, and by the choice of a reflection commuting
	with $G_{+}^{\uparrow}(W_{0})$.

	In many cases, e.g. Minkowski spacetime with Poincar\'e symmetry
	in dimension $d\not=3$, or Minkowski with conformal symmetry in
	any dimension, or de\,Sitter spacetime in dimension $d\not=3$, the
	center of $G_{+}^{\uparrow}(W)$ is one-dimensional, hence
	$\Lambda_{W}$ is fixed up to rescaling.
\end{rem}

Given a (anti)-unitary representation $U$ of $G_{+}$, we can reproduce
the analysis in Section \ref{sec:prim}: Set
$\De_{W}=U(\Lambda_{W}(-i))$, $J_{W}=U(R_{W})$ (the above
normalization at $t=-i$ is conventional, as we could arbitrarily
rescale $\Lambda_{W}$.  The positive energy condition, see below, will
fix the normalization).  Clearly $J_{W}$ is a self-adjoint
antiunitary, and $\De$ is stri\-ct\-ly po\-si\-ti\-ve.  By {\bf (a)}
and {\bf (b)}, $R_{W} = R_{\Lambda_{W}(t)W} =
\Lambda_{W}(t)R_{W}\Lambda_{W}(-t)$, namely $R_{W}$ and $\Lambda_{W}$
commute.  Therefore $J_{W}\De_{W}J_{W}=\De_{W}^{-1}$, and, setting
$S_{W}=J_{W}\De_{W}^{1/2}$, we easily obtain that $S_{W}$ is closed,
densely defined and satisfies $S_{W}^{2}\subset I$.

Set $\calK_{W}=\{\xi\in D(S_{W}):S_{W}\xi=\xi\}$.  It turns out that
$\calK_{W}$ is a standard space, and that the representation $U$ acts
geometrically on the family: $U(g)\calK_{W}=\calK_{gW}$.  Moreover,
essential duality holds: $\calK_{W}'=\calK_{W'}.$

Let $\calB$ be the family of regions that are intersections of wedges,
and set $\calK_{B}=\cap_{W\supset B} \calK_{W}$, $B\in\calB$. 
If we assume $\calW$ to be a subbase for the topology of $M$, then
$\calB$ forms a base, hence any open set $\calO$ is a union of elements
in $\calB$.  Then we may define $\calK_{\calO} =
\vee_{B\subset\calO}\calK_{B}$. $G$-covariance follows as in Section 2.

\begin{Prop}\label{net}
    The following properties hold: 
    \begin{itemize}
    \item[$(i)$]
    $\{\calK_{W},W\in\calW\}$ is a covariant family of real subspaces,
    namely $\calK_{g\calW}=U(g)\calK_{\calW}$, $g\in G_{+}$, moreover
    $\calK_{W}$ is standard and $\calK_{W'}=\calK'_{W}$. 
    \item[$(ii)$] $\{\calK_{B},B\in\calB\}$ is a covariant net of real
    subspaces, namely $B_{1}\subset B_{2}$ implies
    $\calK_{B_{1}}\subset\calK_{B_{2}}$, and
    $\calK_{g\calB}=U(g)\calK_{\calB}$, $g\in G_{+}$.
    \end{itemize}
\end{Prop}

\begin{rem}\label{rem:PCT2}
    As in Remark \ref{rem:PCT}, giving a representation of $G_{+}$ is
    equivalent to giving a representation of $G_{+}^{\uparrow}$
    together with some sort of PCT, namely an anti-unitary involution
    $J$ satisfying $JU(g)J=U(RgR)$, for some reflection $R$.  
\end{rem}

Notice that the net $B\to\calK_{B}$, $B\in\calB$, is not necessarily
local.  Also, it is not necessarily true that $\calK_{B}=\calK_{W}$ if
$B=W$, namely it may happen that $\cap_{W\supset W_{0}}\calK_{W}$ is
strictly smaller than $\calK_{W}$, since we did not prove wedge
isotony.  We need further assumptions to solve these two problems.

Let $H$ be a convex cone in the Lie algebra of $G$, and let us
denote by $\exp$ the exponential map from the Lie algebra to
$G_{+}^{\uparrow}$.  If $W_0\subset W$ are wedges, we shall say that
$W_0$ is positively included in $W$ w.r.t the cone $H$ if there is a
one parameter subgroup $\exp(ah)$ of $G_{+}^{\uparrow}$, depending on
$W_0$ and $W$, with $\exp(a_{0}h) W=W_0$ for some $a_0\geq 0$, such
that $\pm h\in H$, and
\begin{equation*}
\begin{align*}
\Lambda_W(t) \exp(ah) \Lambda_W(-t)&=\exp(e^{\mp 2\pi t}ah)
\\
R_{W}\exp(ah)R_{W}&=\exp(-ah)
\end{align*}
\qquad a,t\in\mathbb R\ .
\end{equation*}
Let us assume the following:
\begin{itemize}
	\item[{\bf (c)}] {\bf Positive inclusion.} Any inclusion of wedges 
	is the composition of finitely many positive inclusions.  
	\item[{\bf (d)}] {\bf Wedge separation.} For any space-like 
	separated $\calO_{1},\calO_{2}\in\calB$ there exists a wedge $W$ 
	such that $\calO_{1}\subset W$ and $\calO_{2}\subset W'$.
\end{itemize}

We shall say that a (anti)-unitary representation of $G_{+}$ is
positive if, whenever $h\in H$, the self-adjoint generators in the
representation space of the one-parameter groups $U(\exp(ah))$ are
positive.

\begin{Thm}\label{genprop}
    Assume the triple $(\calW,R,\Lambda)$ satisfies assumptions {\bf 
    (a), (b), (c), (d)}, and let $U$ be a (anti)-unitary positive
    representation of $G_{+}$.  Then wedge isotony holds,
    namely $W_{1}\subset W_{2}$ implies $\calK_{W_{1}} \subset
    \calK_{W_{2}}$, the net $B\to \calK_{B}$, $B\in\calB$, is local
    and extends the net $W\to\calK_{W}$, $W\in\calW$.  Moreover, for any
    $B\in\calB$ such that $B'\not\in\calB$, Haag duality holds:
    $$
    \calK_{B}'=\calK_{B'}.
    $$
    If $G_{+}^{\uparrow}$ is a simple Lie group with finite center and
    $U$ does not contain the trivial representation, the net is
    irreducible.  If moreover the closure of $\{\Lambda_W(t) :
    t\in\mathbb R\}$ in $G_{+}^{\uparrow}$ is not compact, the local
    space $\calK_{W}$ is a factor.
\end{Thm}

\begin{proof} 
    Wedge isotony follows by property {\bf (c)}, locality of
    $B\mapsto\calK_{B}$ follows by property {\bf (d)}.  If
    $B'_{0}\not\in\calB$ then $\calK_{B'_{0}}=\cup_{B\subset B'_{0}}
    \calK_{B}$, hence Haag duality follows as in Corollary
    \ref{HaagDual}. The assumption of non-compactness for the closure 
    of $\Lambda(t)$ allows us to use the vanishing of the matrix 
    coefficients Theorem as in Theorem \ref{factor} to prove the factoriality.
\end{proof}

Let us observe that, if the positivity in the previous statement is a
non-trivial requirement, namely if there are wedges included one in
another, then $\Lambda$ and $\exp(ah)$ give rise to a representation of
the $ax+b$ group, namely the requirement that the closure of
$\{\Lambda_W(t):t\in\mathbb R\}$ in $G_{+}^{\uparrow}$ is not compact
is automatically satisfied.  For the same reason, also the assumption
on the finiteness of the center is unnecessary (cf. \cite{gl:96}).

Let us discuss a toy example satisfying the general scheme presented 
above, where the last statement of the previous Theorem does not 
apply.  Let $M=S^{2}\times\R$, where $S^{2}$ is the unit sphere in 
$\R^{3}$, with the induced Lorentzian metric, and 
$G_{+}=SO(3)\times\R\rtimes\Ze_{2}$, where $SO(3)$ acts on the sphere, 
$\R$ gives time-translations, and the $\Ze_{2}$ element implements the 
orientation preserving space-time reflection (PT transformation).  We 
also set $\calW$ to be the family of diamonds with base a hemisphere 
(at time $t$).  Clearly the stabilizer $G_{+}^{\uparrow}(W)$ is 
one-dimensional, and, since two hemispheres included one in the other 
coincide, no positivity is needed.  Therefore the parametrization of 
the groups $\Lambda_{W}$ may be fixed arbitrarily.  Also, the action 
of $G_{+}$ is transitive, hence we may fix a wedge $W_{0}$ as the 
causal completion of $\{(t,x,y,z):x^{2}+y^{2}+z^{2}=1,t=0, z>0\}$ and 
assign $$\Lambda_{W_{0}}(\theta)=\left(\begin{matrix} 
1&0&0&0\\0&\cos\theta&\sin\theta&0\\0&-\sin\theta&\cos\theta&0\\0&0&0&1 
\end{matrix}\right),
$$
$R_{W_{0}}(t,x,y,z)=(-t,x,y,-z)$.  In any faithful irreducible
representation of $G_{+}^{\uparrow}$, the generator of $\Lambda_{W_{0}}$ has a
one-dimensional kernel, therefore the corresponding space
$\calK_{W_{0}}$ is not a factor.  More precisely it is a tensor
product of a continuous abelian von Neumann algebra and of a type
$I_{\infty}$ factor (cf. \cite{[FiGu2]}).  However the net is irreducibile.

\medskip

In such a generality, it is not possible to prove important 
properties, such as the standard property, the intersection property, 
or the factor property, for elements of $\calB$.
We now discuss this structure in specific spacetimes.

\subsection{Conformal Group}

In the following the conformal group on the Minkowski spacetime $M$ of 
dimension $d \geq 1$ (with $M=\mathbb R$ if $d=1$) is the group 
generated by the Poincar\'e group (``$ax+b$'' group if $d=1$) and the 
relativistic ray inversion map.  The conformal group is isomorphic to 
$PSO(d,2)$.  If $d>2$, this is the group of local diffeomorphisms 
(defined out of meager sets) which preserve the metric tensor up to 
non-vanishing functions; its universal covering acts globally and 
transitively on the universal covering of the Dirac-Weyl 
compactification of $M$.  If $d=2$, the Dirac-Weyl compactification is 
a two-torus, and only the time-covering is considered, namely the 
conformal group acts on the cylinder spacetime with non compact time 
curves.  In the $d=1$ case the identity component of $PSO(d,2)$ is 
isomorphic to $PSL(2,\mathbb R)$ and we consider its action on $S^1$.  
For details see \cite{bglo1}.

If $d\geq 2$ a wedge is any conformal transformed of (the lift of) a
wedge in the Minkowski space, in particular Poincar\'e-wedges, double cones,
future cones and past cones give rise to conformal wedges.  The maps $R$ and
$\Lambda$ are here the lifts of those defined on Minkowski spacetime. 
If $d=1$ wedges are proper intervals, the reflection associated
with the upper semi-circle maps $z\in S^1$ to its complex conjugate
$\overline{z}$, and $\Lambda$ is the (lift of the) one parameter
subgroup of $PSL(2,\R)$ of (Cayley transformed) dilations.

Then, we may consider (anti-)unitary representations of these groups, 
and check that properties {\bf (a)}, {\bf (b)} and {\bf (c)} hold 
true, the cone $H$ being generated by the Lie algebra generators of 
lightlike translations and their conjugates under the action of the 
conformal group.  In this case wedges form already a base for the 
topology, so it is enough to consider the net on wedges.  Hence 
assumption {\bf (d)} is not needed.  An analogue of 
Theorem~\ref{genprop} holds true here.

\begin{Thm} 
	Let the spacetime be the universal covering of the compactified 
	Minkowski space $S^{d-1}\times\R$ for $d\ge 2$ and $S^1$ if $d=1$.  
	Let $U$ be a (anti-)unitary positive representation of 
	$\widetilde{PSO}(d,2)$ which does not contain the trivial 
	representation.  Then, $\calW\ni W\to\calK_W$ is a local conformal 
	net for which Haag duality holds.  The standard and the III$_{1}$ 
	factor properties are satisfied.  Moreover, the family is 
	irreducible.
\end{Thm}
\begin{proof} 
    Haag duality, conformal covariance and standard property follow by
    Proposition \ref{net} $(i)$, wedge-isotony, factor property and
    irreducibility follow by Theorem \ref{genprop},  and locality 
    follows by wedge isotony and wedge-duality. III$_{1}$ factor 
    property follows as in Proposition 1.2, \cite{gl:96}.
\end{proof} 

The one-dimensional conformal case is extensively studied in
\cite{GLW} and we refer to that paper for further details.  We recall
that all the nets corresponding to irreducible representations of
$PSL(2,\R)$ are subsystems ($n$-th derivatives) of the same net on
$\mathbb R$ (the $U(1)$ current algebra) which is their common dual
net.

\subsection{de Sitter spacetime}

Since the $d$-dimensional de\,Sitter spacetime $dS^d$ may be defined as 
the hyperboloid $x_0^2+1=\sum_{i=1}^d x_i^2$ in $M^{d+1}$, the wedges can 
be defined as the intersection of this hyperboloid with the wedges in 
$M^{d+1}$ whose edges contains the origin.

The natural symmetry group of $dS^{d}$ is the Lorentz group
$\calL_{+}=SO(d,1)$, and the maps $R$, $\Lambda$ are assigned here as
in the Minkowski spacetime.  Then properties {\bf (a)}, {\bf (b)}, and
{\bf (d)} immediately follow from the corresponding properties for
the Minkowski spacetime.  Property {\bf (c)} instead is trivially
satisfied, since two wedges $W_{1}\subset W_{2}$, whose edge contain
the origin, coincide.

Intersections of wedges, namely elements of $\calB$, correspond to 
spacelike cones in the Minkowski space, therefore the standard 
property and the intersection property on the $d$-dimensional de\,Sitter 
spacetime can be studied applying the techniques of the preceding 
section. But we can also rely on the direct analysis by Bros and 
Moschella \cite{BrMo}.

Let us recall that the irreducible representations of the group
$SO(d,1)$, $d\geq2$, belong to three classes, usually called principal
series representations, complementary series representations, and
discrete series representations (cf.  e.g.
\cite{Vilenkin,Thieleker1}).  The first class corresponds to
representations appearing in the direct integral decomposition of the
regular representation, the second one to representations not
appearing in the direct integral decomposition of the regular
representation.  Concerning the third class however, the name
``discrete series'' is not always appropriate, namely it is not always
true that they are irreducible direct summands of the regular
representation.  Indeed, applying a result of Harish-Chandra, it turns
out that this fact is possible if and only if $d$ is even (see
\cite{Thieleker1}).  The exact determination of the direct-summand
representations, namely the recognition of the discrete series as
opposed to the ``mock discrete series'' is well known for the
two-covering $SL(2,\R)$ of $SO(2,1)$ \cite{[L]}, implying that there
are not mock discrete series representations for $SO(2,1)$.  This
problem has been solved in \cite{Dix} for $d=4$ and in
\cite{Thieleker2} for a general $d=2m$, $m\geq2$.

\begin{Thm}
	Let $\{\calK(B):B\in\calB\}$ be the net of local real vector 
	subspaces associated to a representation $U$ of the Lorentz group 
	$SO(d,1)$.  If $U$ is a subrepresentation of the regular 
	representation, then the standard property, the intersection 
	property and the factor property hold.  If $U$ is a representation 
	in the principal or complementary series of the Lorentz group 
	$SO(d,1)$, then the mentioned properties hold.
\end{Thm}

\begin{proof}
	The restriction of a representation from the Poincar\'e group to 
	the Lorentz group gives a map from nets $\calK^{M}$ on the 
	Minkowski space $M^{d}$ to nets $\calK^{S}$ on the de\,Sitter 
	space defined as $\calK^{S}(B)=\calK^{M}(\calC(B))$, where 
	$\calC(B)$ is the spacelike cone in $M^{d}$ generated by the 
	region $B$ in $S^{d-1}$ and the origin.  We may rephrase results in 
	the previous section saying that the standard property, the 
	intersection property and the factor property hold for regions $B$ 
	given as intersections of wedges in the regular representation of the 
	Lorentz group.
	
	By Theorem \ref{mgives0}, cf.  also Remark \ref{higherdim},
	the properties hold for all subrepresentations.  Since the
	regular representation decomposes as direct integral of the
	principal series representations, the standard and the
	intersection properties hold for almost all values of the
	parameter labeling the principal series.
	
	However, we may use the analysis in \cite{BrMo}, where it is
	shown that a class of free fields may be constructed,
	corresponding to the principal, resp.  complementary series of
	the representations of $\Lor$.  In \cite{BrMo} the authors
	prove the Reeh-Schlieder and Bisognano-Wichmann properties for
	the free fields corresponding to the principal series, and
	state that these results extend to the complementary series. 
	By the Bisognano-Wichmann property, such free fields
	necessarily give rise to the nets constructed as above for the
	corresponding representations.  Therefore the standard property
	follows by the Reeh-Schlieder property and the intersection
	property is trivially satisfied since the local algebras are
	generated by local fields.
\end{proof}

Thus, concerning the principal and complementary series, the standard
and intersection properties are consequence of the Reeh-Schlieder in
\cite{BrMo}.  Yet the above proof goes beyond that, by showing the
same properties to hold in the $dS^{d}$ models associated with the
discrete series, $d$ even.  Discrete series representations have been
explicitly excluded in the analysis in \cite{BrMo}, and the result
that such representations give rise to (free) nets of local algebras
on the de Sitter space-time seems to be not known before our analysis. 
This will be discussed in detail in \cite{inp}.

\appendix

\section{Restricting the Poincar\'e group representations to the
Lorentz subgroup}

We give here an analysis of the representations of the Lorentz and
Poincar\'e groups needed in the paper.  We treat explicitly the
$3+1$-dimensional case, however the analysis extends to any dimension,
as explained in Remark \ref{higherdim}.

If $G$ is a locally compact group, we shall denote by $\lambda_G$ its 
left regular representation.  If $H\subset G$ is a closed subgroup and 
$\pi$ is a unitary representation of $H$, we shall denote by 
$\Ind_{H\uparrow G}(\pi)$ the representation of $G$ induced by $\pi$ 
in the sense of Frobenius, Wigner and Mackey; we shall refer to the 
books \cite{[Z],[K],Lip} for the theory of induced representations.

Let us recall that the irreducible representations of $\Poif$ are
induced representations $\Ind_{F\uparrow \Poif}(\eta)$, where $F=F(p)$
is the stabilizer of some point $p\in\R^{4}$, $\Poif$ acting on the
subgroup $\R^{4}$ by conjugation, and $\eta$ is an irreducible
representation of $F$.  We often identify $\R^4$ with its dual
$\hat\R^4$.  When $p$ varies in a given $\Poif$-orbit the
corresponding induced representations are equivalent, therefore they
are labelled by $m=p^{\mu}p_{\mu}$.  When $m>0$ the stabilizer is
isomorphic to $SO(3)\ltimes\R^{4}$, therefore positive mass $m$ representations are
completely described by the spin $s$.  When $m=0$ and we choose $p_{0}>0$
(to have positive energy), the stabilizer is isomorphic to the
Euclidean group $E(2)$.  The representations which are trivial on the
$E(2)$-translations are the so-called finite-helicity representations,
and are completely labelled by $\Ze$.  The others are called
continuous-spin representations.

The other cases, namely $p=0$ and $m<0$, correspond respectively to
null energy (trivial translations) and non positive energy.

In the following we shall say that a property $P$ for representations 
of a group $G$ is {\sl stable} if ``$P$ is true for $\pi$'' implies  
``$P$ is true for all representations unitarily equivalent to $\pi$'' and
\begin{equation}
P\ \text{is true for}\ \pi\equiv \pi_1\oplus\pi_2\Leftrightarrow
P\ \text{is true for}\ \pi_1 \ {\rm and}\ \pi_2 \, .
\end{equation} 
\begin{Thm}\label{mgives0} 
	Assume that $P$ is a stable property for the representations of $\Lor$.
	The following are equivalent:
	\begin{itemize}
\item[$(i)$] $P$ is true for the restriction to $\Lor$ of the positive mass 
	representations.
\item[$(ii)$] $P$ is true for the restriction to $\Lor$ of 
	the continuous spin representations.
\item[$(iii)$] $P$ is true for the restriction to $\Lor$ of 
the massless finite helicity representations.
\item[$(iv)$] $P$ is true for $\lambda_{\Lor}$.
\end{itemize}
	\end{Thm}

The proof of this theorem requires some steps.

\begin{Lemma}\label{restrind}
	Let $\pi=\Ind_{F\uparrow \Poif}(\eta)$ be an irreducible representation 
	of $\Poif$ as above, $F=F(p_{0})$. Then 
	$$
	\pi|_{\Lor}=\Ind_{E\uparrow \Lor}(\eta|_E),
	$$
	where $E$ is the stabilizer, in $\Lor$, of $p_{0}$.
\end{Lemma}

\begin{proof}
	Let us denote by $X$ the orbit of $p_{0}$ under the adjoint action 
	of $\Poif$ on the subgroup $\R^{4}$, or equivalently the 
	homogeneous space $\ProPoi/F(p_{0})$, and let $\nu$ be the
	$\Poif$-invariant measure on $X$.  If $\eta$ is a 
	representation of $F$ acting on the Hilbert space $\calH_{\eta}$ 
	of $\eta$, $\pi$ can be defined as
	$$
	(\pi(g)\xi)(p)=\eta(\al(g,p))\xi(g^{-1}p),
	$$
	where $g\in\Poif$, $p\in X$, $\xi\in L^{2} (X, 
	\calH_{\eta}, d\nu)$, and $\al$ is an $F$-valued cocycle of the form 
	$\al(g,p)=s(p)^{-1}gs(g^{-1}p)$, where $s$ is a 
	Borel section, namely a Borel map $s:X\to\Poif$ satisfying $s(p)p_{0}=p$.  
	
	By definition, $\R^{4}$ acts trivially on itself, hence 
	$F=E\ltimes\R^{4}$ with $\R^{4}$ acting trivially on $X$, 
	therefore we may choose $s$ to be a $\Lor$-valued section.  As a 
	consequence, $\al:\Lor\times X$ is $E$-valued, namely the 
	restriction to $\Lor$ of $\pi$ is by definition the representation 
	induced by $\eta|_E$.
\end{proof}

If $\rho$ and $\sigma$ are representations, we shall write 
$\rho=\sigma$ if $\rho$ is unitary equivalent to $\sigma$ and
$\rho\approx\sigma$ if $\rho$ is quasi equivalent to $\sigma$, namely
$\rho\otimes\iota=\sigma\otimes\iota$, where $\iota$ is the identity 
representation on $\ell^2(\mathbb N)$.

\begin{Lemma}\label{HG2}
	Let $H$ be a locally compact group isomorphic to the Euclidean 
	group $E(2)$.  If $\pi$ is an irreducible unitary representation 
	of $H$ and $\pi$ has non-trivial restriction to the subgroup 
	$\mathbb R^2$, then $\pi=\pi_{q}\equiv\Ind_{\mathbb R^2\uparrow 
	H}(q)$ where $q\neq 0$ is a character $q\in\hat\mathbb R^2$.

	We have $\lambda_H=\int^{\oplus}_{\hat\mathbb R^2} 
	\pi_{q}\text{d}q$.
\end{Lemma}
\begin{proof}
$E(2)$ is the semidirect product $E(2)=\mathbb R^2\rtimes \mathbb T$, 
where $\mathbb T$ acts on the plane $\mathbb R^2$ by rotations. The 
action of $E(2)$ on $\hat\mathbb R^2$ by dual conjugation factors 
through the action of $\mathbb T$ and is smooth. 
The stabilizer $H_q$ of a point $q\in\hat\mathbb R^2$ is  
$E(2)$ (iff $q=0$) or $H_q=\mathbb R^2$.
By Mackey's theorem every irreducible representation $\pi$ of $H$ is 
induced from an irreducible representation $\rho$ of $H_q$ with
$\rho |_{\mathbb R^2} = \text{dim}(\rho)q$.

Thus either $q=0$ and $\pi$ acts trivially on $\mathbb R^2$, or 
$q\neq 0$ and $\pi = \pi_q$.

The rest is now clear by induction at stages because
\[
\lambda_H=\Ind_{\mathbb R^2\uparrow H}(\lambda_{\mathbb R^2})
=\Ind_{\mathbb R^2\uparrow H}(\int^{\oplus}_{\hat\mathbb R^2} 
q\text{d}q)
=\int^{\oplus}_{\hat\mathbb R^2}\Ind_{\mathbb R^2\uparrow H}(q)\text{d}q=
\int^{\oplus}_{\hat \mathbb R^2}\pi_{q}\text{d}q .
\]
\end{proof}

\begin{Prop}\label{reg}
 Let $G$ be a locally compact group and $H\subset G$ a 
closed subgroup isomorphic to the Euclidean group $E(2)$. Then
\[
\lambda_G =\int^{\oplus}_{\hat\mathbb R^2}\Ind_{H\uparrow 
G}(\pi_q)\text{d}q .
\]
\end{Prop}
\begin{proof} Immediate by the Lemma \ref{HG2} because
\[
\lambda_G =\Ind_{H\uparrow G}(\lambda_H)=
\Ind_{H\uparrow 
G}(\int^{\oplus}_{\hat\mathbb R^2}\pi_q\text{d}q)
=\int^{\oplus}_{\hat\mathbb R^2}\Ind_{H\uparrow 
G}(\pi_q)\text{d}q .
\]
\end{proof}
We shall denote by $\pi_{m,s}$ the irreducible representation of 
mass $m>0$ and spin $s\in\mathbb N$ of the Poincar\'e 
group $\Poif$ and by $\pi^{\Lor}_{m,s}$ its restriction to the Lorentz 
subgroup $\Lor$.

 By definition, the continuous spin representations $\si_{q}$ of $\Poif$ are 
 the ones induced by the representations $\pi_{q}$ of 
 $H=E(2)\ltimes\R^4$ in Lemma \ref{HG2}, where $H$ is the stabilizer 
 in $\Poif$ of a point $p$ with
 $\langle p,p\rangle =0$, $p_0 > 0$.  We shall 
 denote by $\si_q^{\Lor}$ the restriction of $\si_{q}$ to $\Lor$.  
 By Lemma \ref{restrind} we have $\si_q^{\Lor}=\Ind_{\mathbb R^{2}\uparrow\Lor}(q)$.
\begin{Lemma}\label{decomp}
$\lambda_{\Lor}=\int^{\oplus}_{\hat\mathbb R^2} \si_q^{\Lor}\text{d}q.$
\end{Lemma}
\begin{proof} 
Immediate by Lemmas \ref{restrind}, \ref{HG2} and Prop. \ref{reg}.
\end{proof}
\begin{Lemma}\label{m>0}
For any given $m>0$ we have
\[
\lambda_{\Lor}\approx\bigoplus_{s\in\mathbb N}\pi^{\Lor}_{m,s} .
\]
\end{Lemma}
\begin{proof} Denote by $\rho_s$ the representation of $SO(3)$ of spin 
$s$.  By Lemma \ref{restrind} $\pi^{\Lor}_{m,s} \approx 
\Ind_{SO(3)\uparrow\Lor} (\rho_s)$, in particular $\pi^{\Lor}_{m,s}$ 
is independent of $m>0$.

We have
\[
\bigoplus_{s\in\mathbb N}\pi^{\Lor}_{m,s}=
\bigoplus_{s\in\mathbb N}\Ind_{SO(3)\uparrow\Lor}(\rho_s)=
\Ind_{SO(3)\uparrow\Lor}(\bigoplus_{s\in\mathbb N}\rho_s)\approx
\Ind_{SO(3)\uparrow\Lor}(\lambda_{SO(3)})\approx\lambda_{\Lor} .
\]
\end{proof}
The following Lemma is a particular case 
of the subgroup Theorem of Mackey (cf. e.g. \cite{Lip}, 
Chapter II, Theorem 1) when the subgroup $G_{2}$ coincides with the 
group $G$. We give a proof here for the convenience of the reader.

\begin{Lemma}\label{trivMackey}
	Let $H$ be a subgroup of $G$, $\eta$ a representation of $H$ and $g_0$ 
	an element of $G$ normalizing $H$. Then 
	$$
	\Ind_{H\uparrow G}(\eta)= \Ind_{H\uparrow G}(\eta^{g_0}),
	$$
	where $\pi^{g_{0}}(h)\equiv\pi(g_0^{-1}hg_0)$, $h\in H$.
\end{Lemma}
\begin{proof}
	Let us denote by $X$ the homogeneous space $G/H$, and let $\nu$ be 
	a $G$-quasi-invariant measure on $X$, that for simplicity we 
	assume to be invariant.  Setting $\pi=\Ind_{H\uparrow G}(\eta)$, 
	namely
	$$
	(\pi(g)\xi)(p)=\eta(\al(g,p))\xi(g^{-1}p),
	$$
	where $g\in G$, $p\in X$, $\xi\in L^{2} (X, 
	\calH_{\eta}, d\nu)$, and $\al$ is an $H$-valued cocycle
	$\al(g,p)=s(p)^{-1}gs(g^{-1}p)$, with $s:X\to G$  a 
	Borel section satisfying $s(p)p_{0}=p$. 

Hence $\pi^{g_0}$ is given by
\begin{multline}
	(\pi^{g_0}(g)\xi)(p)=\eta(g_0^{-1}\al(g,p)g_0)\xi(g^{-1}p)\\
	=\eta(g_0^{-1}s(p)^{-1}gs(g^{-1}p)g_0)\xi(g^{-1}p)
	=\eta(\al^{g_0}(g,p))\xi(g^{-1}p),
\end{multline}
where the cocycle $\al^{g_0}(g,p)=g_0^{-1}s(p)^{-1}gs(g^{-1}p)g_0=
s^{g_0}(p)^{-1}gs^{g_0}(g^{-1}p)$ is associated with the map
$s^{g_0}(p)=s(p)g_0$. Clearly $s^{g_0}:X\to G$ is a Borel section 
for the different quotient map $g\to gg_0^{-1} p_0$. As the stabilizer of 
$p_0$ coincides with the stabilizer of $g_0^{-1} p_0$, the statement 
follows by the uniqueness of the induced representation.
\end{proof}

\begin{proof} {\it(of Theorem \ref{mgives0})} $(i)\Leftrightarrow (iv)$: 
By Lemma \ref{m>0}, property $P$ holds 
	for the representations $\pi^{\Lor}_{m,s}$ iff it holds 
	for the regular representation, by stability.
	
	$(ii)\Leftrightarrow (iv)$:
	If $p\ne0$ has zero mass, the stabilizer $E(p)$ in $\Lor$ does not 
	change replacing $p$ with $\lambda p$, $\lambda>0$. Therefore all 
	elements in $\Lor$ moving $p$ to some of its multiples normalizes the 
	stabilizer of $p$. For such a $g$, 
	$$
	\pi_q(g^{-1}hg)=\pi_{gq}(h), \ h\in E(p).
	$$
	Note that every $p$-orbit in $\hat{\R}^{2}$ (except $\{0\}$) can be 
	reached by some $g$ with the property $g:p\mapsto\lambda p$.
	Therefore the $\sigma^{\Lor}_{q}$'s are all equivalent, by Lemma 
	\ref{trivMackey}. Then, by Lemma \ref{decomp}, $\sigma^{\Lor}_{q}$ 
	is a subrepresentation of the regular representation, hence 
	property $P$ holds by stability. The converse is also true by 
	stability.
	
	$(iii)\Leftrightarrow (iv)$: The argument is again similar to the 
	above ones. Let $\chi_n\in\hat\mathbb T$, $n\in\mathbb Z$ be the 
	characters of $\mathbb T$. The finite helicity representations are the 
	representations of the Poincar\'e group
	induced by the representations $\alpha_n\equiv\Ind_{\mathbb T\uparrow 
	E(2)}(\chi_n)$ of $E(2)$. Their restrictions to $\Lor$ are
	$\Ind_{E(2)\uparrow\Lor}(\alpha_n)$. Then, by induction at stages,
	\[
	\bigoplus_n \Ind_{E(2)\uparrow\Lor}(\alpha_n)=
	\bigoplus_n \Ind_{\mathbb T\uparrow\Lor}(\chi_n)=
	\Ind_{\mathbb T\uparrow\Lor}(\bigoplus_n \chi_n)=
	\Ind_{\mathbb T\uparrow\Lor}(\lambda_{\mathbb T})=\lambda_{\Lor},
	\]
	and the statement follows by stability.
\end{proof}

\begin{rem}\label{higherdim}
    Although the proof of Theorem \ref{mgives0} has been written
    for the 4-dimensional case, it extends to the case of the
    Poincar\'e group $\Poif(d)$ acting on the $d$-dimensional
    Minkowski space, $d\geq2$.  Indeed the continuous spin
    representations are present only when $d\geq4$, therefore the
    property $(ii)$ is void for dimension $\leq3$.  When $d\geq4$,
    the stabilizer of a light-like point is the Euclidean group
    $E(d-2)$, whose irreducible representations are parametrized by
    vectors in $\R^{d-2}$ (and vectors with the same length give
    equivalent representations).  This can be found e.g. in
    \cite{Vilenkin}, or proved by induction where the first is
    given by Lemma \ref{HG2} and the induction step follows by the
    Mackey Theorem (\cite{[Z]}, Thm 7.3.1).  Therefore all the above
    analysis applies.
\end{rem}

\end{document}